# Network and Station-Level Bike-Sharing System Prediction: A San Francisco Bay Area Case Study

Huthaifa I. Ashqar, Ph.D. (Corresponding author)
Booz Allen Hamilton
Washington, D.C. 20003, United States
hiashqar@vt.edu

Mohammed Elhenawy, Ph.D.
CARRS-Q, Queensland University of Technology
Kelvin Grove QLD 4059, Australia
mohammed.elhenawy@qut.edu.au

Hesham A. Rakha, Ph.D., P.Eng.
Charles E. Via, Jr. Department of Civil and Environmental Engineering
Virginia Tech Transportation Institute
3500 Transportation Research Plaza, Blacksburg, VA 24061, United States
hrakha@vt.edu

Mohammed Almannaa, Ph.D.
Charles E. Via, Jr. Department of Civil and Environmental Engineering
Virginia Tech Transportation Institute
3500 Transportation Research Plaza, Blacksburg, VA 24061, United States
malmannaa@ksu.edu.sa

Leanna House, Ph.D.
Department of Statistics, Virginia Tech
Blacksburg, VA 24061, United States
lhouse@vt.edu

Word count: 5,851 words text + 8 tables/figures

# Network and Station-Level Bike-Sharing System Prediction: A San Francisco Bay Area Case Study

## ABSTRACT

The paper develops models for modeling the availability of bikes in the San Francisco Bay Area Bike Share System applying machine learning at two levels: network and station. Investigating BSSs at the station-level is the full problem that would provide policymakers, planners, and operators with the needed level of details to make important choices and conclusions. We used Random Forest and Least-Squares Boosting as univariate regression algorithms to model the number of available bikes at the station-level. For the multivariate regression, we applied Partial Least-Squares Regression (PLSR) to reduce the needed prediction models and reproduce the spatiotemporal interactions in different stations in the system at the network-level. Although prediction errors were slightly lower in the case of univariate models, we found that the multivariate model results were promising for the network-level prediction, especially in systems where there is a relatively large number of stations that are spatially correlated. Moreover, results of the station-level analysis suggested that demographic information and other environmental variables were significant factors to model bikes in BSSs. We also demonstrated that the available bikes modeled at the station-level at time $t$ had a notable influence on the bike count models. Station neighbors and prediction horizon times were found to be significant predictors, with 15 minutes being the most effective prediction horizon time.

## Introduction

In the next few decades, many traditional cities will be turned into smart cities, which are greener, safer, and faster. This transformation is supported by recent advances in information and communication technology (ICT) in addition to the expected fast spread of the Internet of Things (IoT) and big data analytics. Smart cities may mitigate some of the negative impacts of traditional cities. Smart cities have many components, including smart transportation. Smart transportation will integrate different transportation networks and allow them to work together so travelers and commuters can enjoy seamless multi-modal trips based on their preferences. Consequently, more commuters will be inspired to use public transportation systems and many traffic-related problems such as congestion could be reduced [1-3].

Using bike-sharing systems (BSSs) seems to be a promising potential system to decrease pollution, traffic congestion, and gas emissions [2]. Recently, BSSs are gaining more attention as a crucial portion of micromobility urban approach in different cities because they seem to be sustainable and environmentally friendly. While the first established BSS in the United States was in Portland in 1964, many cities nationally such as Washington D.C. and San Francisco have also found BSS programs implementing various structures, goals, conditions, and strategies. In this paper we used a data from the Bay Area Bike Share System (BSS) (now called the "Ford GoBike" BSS), which was found in 2013 in San Francisco as a membership-based system. The system provides a self-service access as a short-term rental bike system during anytime. The idea is very basic, a member will depart from a station in the network, ride to the nearest station to the destination, and lock the bicycle safely in that station leaving it to someone else to use [4]. Most of the trips in the Bay Area BSS is short and quick trips, as there is an additional fee for trips that takes more than 30 minutes. The system, which included 70 stations in 2013 to 2015 connects four different areas: Palo Alto, downtown San Jose, downtown San Francisco, and Mountain View [4]. The system is being used by daily commuters and tourists who would like to get across these different areas at any time but especially during the rush hour [4].

In order to facilitate the forecasting of the fluctuated bike counts at each station in BSSs, this study used three different machine learning algorithms to predict the available bikes at a bike-sharing station in the Bay Area Bike Share System. This will help operators and policymakers to coordinate and take decisions in such a large, complicated, polluting, and expensive problem [5]. We proposed using Random Forest (RF) and Least-Squares Boosting (LSBoost) algorithms to model univariate algorithm to predict the number of available bikes at each Bay Area bike station. As will be shown in this study and previous studies, a station-level analysis is an important key task to make operating BSSs more efficient and less expensive. However, we also used Partial Least-Squares Regression (PLSR), which is a multivariate model, to build the least required but sufficient prediction models for the network-level analysis.

## Related work

Predicting the number of bike available in BSSs has gain attraction of researcher lately. Previous studies have investigated the effect of many factors such as time, day, month, weather variables, environment indicators, and transportation infrastructure. These studies were implemented for different goals, which include facilitate the rebalance process [6-9], to obtain new real world insights on the interactions between bike trips and other features [10-14], and, ultimately, help policymakers and operators in optimizing their decisions [10, 15-18].

Many studies have been performed at the station-level to predict the availability of bikes by using time series analysis. Rixey applied a multivariate linear model to study station-level BSS ridership, investigating its relationship with different factors including job density, population density, number of commuters using different modes, median income, bikeways infrastructure, education, population race, precipitation days, and the vicinity to other transportation networks' stations [13]. The author found that demographics information, environmental factors, and the closeness to complementary transportation network's stations were essential factors in predicting ridership.

Many researchers have used time series analysis. For example, Froehlich, Neumann, and Oliver applied four different models to forecast the available bikes at each station, namely, the last value, the historical mean, the historical trend, and the Bayesian networks [19]. Furthermore, autoregressive-moving average (ARMA) and autoregressive integrated moving average (ARIMA) were other models of time series analysis that have also been applied. Kaltenbrunner *et al.* used ARMA [20] while Yoon *et al.* modified ARIMA model to take into account the spatiotemporal correlation [21]. Moreover, applying a seasonal ARIMA modeling and considering the complex serial correlation patterns in the error terms with the examining of the model against real bicycle counts, Gallop *et al.* exploiting hourly bike counts and other weather factors to predict bike traffic in Canada [22]. Results showed that weather is a significant factor that affect bike usage such as temperature, rain, humidity, and clearness.

However, few studies have used machine learning to model bike sharing data. One of the characteristics of transportation-related datasets is that they are usually very big – *Big Data*. As a result, applying machine learning algorithms to identify potential explanatory factors is considered promising and robust [15]. To avoid overfitting, different cross-validated algorithms have been used to predict bike availability in a BSS, such as support vector machine (SVM), random forest (RF), and gradient boosted tree (GBT) [23-28]. The authors of the four studies in [23, 24, 26, 27] used different machine learning algorithms to predict bike demand based on the usage record and other information about the targeting prediction time window. While the full prediction problem would be predicting bike counts at each station, the authors used machine learning to predict the bike count of the entire BSS instead. In [25], the authors used RF to classify the stations only with regard to whether the station was full of bikes or totally empty, so users could not return a bicycle, or could not find one to rent. In [29], the authors used PLSR to propose an "extended norms activation model (NAM) to study the effects of personal norms, perceived green value and perceived pleasure on users' green intention [but in a dockless bike-sharing system analyzing] data collected from 308 participants in Chengdu China."

However, this study proposes three main contributions to the literature. 1) modeling bike count prediction at the station-level using machine learning algorithms has not been studied well to date. 2) to consider the spatiotemporal correlation between different stations and to facilitate implementing this when applied to relatively large BSS network. And 3) Station neighbors are taking into consideration as a potential predictor of the regression models but are identified by the trip's adjacency matrix.

## METHODS

### Random Forest Algorithm (RF)

RF was proposed by Breiman as a supervised machine learning technique that could be used for classification and regression purposes [30]. The way that RF works is by creating an ensemble of many decision trees and then it selects a random subset of variables to produce each tree in the ensemble. As each tree is being trained, RF employs a criterion in several steps (i.e. nodes) to divide the trained data. The forest rate error depends on two factors the correlation between two different trees and the significance of each tree as an individual in the forest. Practically, RF uses the mean squared error of the responses from each tree for regression.

In this study, we used RF as it offers many pros [30, 31]. Some of these related advantages are there are very few or no assumptions when applying it in most of the cases, it is well-suited and robust when there is a risk of overfitting, there are no need to search for data transformations to fit, it is robust against multicollinearity (can handle extremely linear and nonlinear relationships between continuous numerical variables and categorical), it can find the relative *importance* of each variable by ranking their contribution in the predictive model.

### Least-Squares Boosting Algorithm (LSBoost)

LSBoost is a gradient boosting of regression trees that produces highly robust and interpretable procedures for regression. LSBoost was proposed by Friedman as a gradient-based boosting strategy [32], using square loss $L(y, F) = (y - F)^2/2$, where $F$ is the actual training and $y$ is the current cumulative output $y_i = \beta_0 + \sum_{j=1}^{i-1} \beta_j h_j + \beta_i h_i = y_{i-1} + \beta_i h_i$. The new added training $\hat{F}$ is set to minimize the loss, in which the training error is computed as in [33]:

$$E = \sum_{t=1}^{N} \left[\beta_i h_i^t - \hat{F}^t\right] \qquad (1)$$

where $\hat{F}$ is the current residual error and the combination coefficients $\beta_i$ are determined by solving $\partial E/\partial \beta_i = 0$.

We used RF and LSBoost were applied in the station-level analysis to predict the number of available bikes in each station at any time $t$. Although both RF and LSBoost are ensemble learning machine learning algorithms that grow decision trees to build robust models, they differ in the way that each component tree is trained in these algorithms. While RF uses randomness and grow each decision tree independently, LSBoost tend to grow only one tree at a given time and adds new tree to its structure by correcting errors resulted from the pre-trained trees.

### Partial Least-Squares Regression Algorithm (PLSR)

PLSR was proposed as a multivariate response algorithm and developed recently in [34-38]. The way that PLSR works is by looking for a linear regression model to project the response variables ($Y$) and the covariate features ($X$) to a new dimensional space. The essential building block of the PLSR method includes implementing a regression between two blocks namely $X$ and $Y$. PLSR then creates different *outer* relations for each of the $X$ and $Y$ blocks in the new space, and another *inner* relation in the purpose of correlate both blocks. PLSR offers many advantages making it the best model for our data. It can robustly handle various descriptor variables, as well as nonorthogonal variables that may need multiple responses, while at the same time it still can provide a promising predictive accuracy and it reduces the risk of correlation.

## DATASET

This study utilized two related anonymized bike trip datasets that was gathered from August 2013 to August 2015 in the San Francisco Bay Area, which is shown in Fig. 1 [39]. Station ID, corresponding number of bikes available and docks available, and the time of recording an incident was extracted from the first dataset. Specifically, from the time data, we extracted the year, month, day-of-month, day-of-week, time-of-day, and the minute when an incident is recorded. This dataset contains a high number of documented events, as we found that it documented every incident occurred per minute for 70 stations in the system over the two years of the period of the study. Consequently, and to relax the complexity, the number of observations was reduced by selecting incidents for every 15 minutes but by extracting the exact values without the need for tuning process.

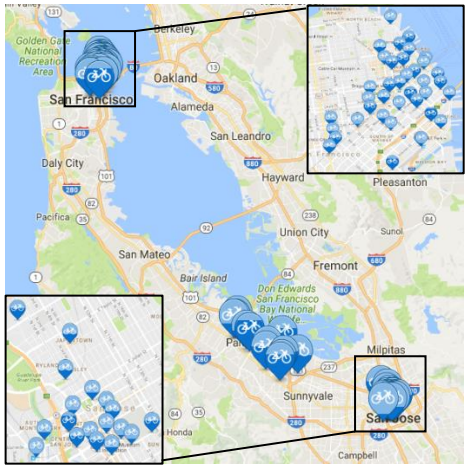

Fig. 1. Stations map. (Source: Google Maps)

As we pointed out in a related study [28], "during the data processing phase, we found that numerous stations had recently been added to the network and others had been terminated. As a result, the dataset was cleaned by eliminating any entries missing docking station data. This reduced the number of entries from approximately 70,000 to 48,000. Each entry included the availability of bikes at the 70 stations with the associated time (month of year, day of week and hour of day) and the weather information". The weather information included many features, but we only chose to use temperature, visibility, humidity, wind speed, precipitation level, and if it was a rainy, foggy, or sunny day. These parameters were selected based on subject-matter expertise and previous related studies [12, 22], and were found to be significant in predicting the number of available bikes at Bay Area BSS stations [40-44].

Moreover, trip data included detailed information about origin station, destination station, and time of each bike trip within the BSS during the 2 years. We used the trip data to generate the BSS network adjacency matrix and we concluded that the highest 10 in-degree-stations for a station $i$ could be allocated as *neighbors* of station $i$. In other words, the neighbors of a station $i$ were defined based on the number of trips that originated from station $j$, in which $j \neq i$, and ended at station $i$.

RESULTS

**Univariate Models**

RF and LSBoost algorithms were applied to create univariate models to predict the number of available bikes at each of the 70 stations of the Bay Area BSS network. The two algorithms were applied to investigate the effect of several variables on the prediction of the number of available bikes in each station $i$ in the network, including the available bikes at station $i$ at time $t$, the available bikes at its neighbors at the same time $t$, the month of year, day of week, time of day, and various selected weather conditions. The predictors' vector for station $i$ at time $t$, denoted by $X_t^i$, was used in the built models to predict the $log$ of the number of available bikes at station $i$ at time $t + \Delta$, which is denoted by $\log(y_{t+\Delta}^i)$, where $i = 1, 2, \ldots, 70$ and $\Delta$ is the prediction horizon time. The effect of different prediction horizons, $\Delta$ (range 15–120 minutes), on the performance of both algorithms was investigated by finding the Mean Absolute Error (MAE) per station (i.e., bikes/station), which can be described as the prediction error. Moreover, as the number of generated trees by RF and LSBoost is an important parameter in implementing both algorithms, we investigated its effect by changing the number of generated trees from 20 to 180 with a 40-tree step.

As shown in Fig. 2 and Fig. 3, the prediction errors of RF and LSBoost increased as the prediction horizon $\Delta$ increased. The lowest prediction error for both algorithms occurred at a 15-minute prediction horizon. Moreover, the prediction error of RF and LSBoost decreased as the number of trees increased until it reached a point where increasing the number of trees would not significantly improve the prediction accuracy. Fig. 2 and Fig. 3 also show that a model consisting of 140 trees yields a relatively sufficient accuracy.

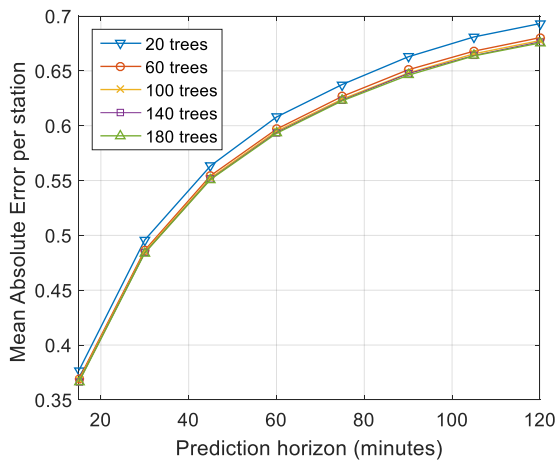

Fig. 2. RF MAE at different prediction horizons and number of trees.

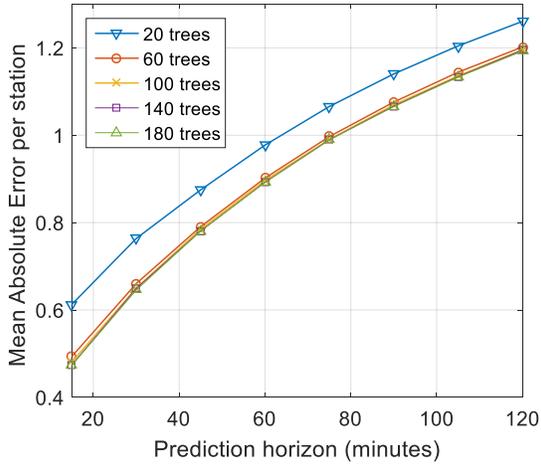

Fig. 3.  LSBoost MAE at different prediction horizons and number of trees.

**Station-Level Analysis**

Investigating the BSS at the station-level is the comprehensive prediction issue that would help planners, operators, and policymakers make effective decisions such as considering the relocation of underused stations (or building new ones) to serve busier region in the bike-sharing network. In order make a station-level analysis of the BSS, we used RF, which consisted of 140 trees at a prediction horizon time of 15 minutes, to model the bike availability at each bike station in the BSS network. The behavior of the algorithm at different stations was investigated by finding the maximum absolute error ($MaxAE = max(|y_{t+\Delta}^i - \hat{y}_{t+\Delta}^i|)$ where $i = 1, 2, ..., 70,$ and $\Delta=$ 15 minutes) at each station as shown in Fig. 4. We used $MaxAE$ to explore the max prediction error that might occur in some time $t$ at each station using 5-fold of cross validation. We used $MaxAE$ to explore the max prediction error that might occur in some time t at each station using 5-fold of cross validation. As a result, 80% of the cells of the training set was split up into training and the rest was considered for testing, which is about 20% of the cells. Cross-validation is a technique for "assessing how the results of a statistical analysis will generalize to an independent data set, which means it estimates how accurately a predictive model will perform in practice". Cross-validation aims "to test the model's ability to predict new data that was not used in estimating it, in order to flag problems like overfitting or selection bias and to give an insight on how the model will generalize to an independent dataset" [45].

Fig. 4. Illustrates that predicting bikes in some stations, specifically the first 32 stations and the last two stations, produced relatively lower $MaxAE$ than the other stations. Looking for those stations, we found that the stations that produced higher $MaxAE$ than the others are the downtown San Francisco's (SF) stations. SF is known to have a high population density compared to the other three regions that the BSS extends to [46]. Moreover, we hypothesized that there is a relation between the users of public transportation and the users of BSS in SF. As the annual report of TomTom Traffic Index 2017 [47] shows that "drivers in San Francisco expecting to spend an average of 39% extra travel time stuck in traffic anytime of the day, which is 7% more than San Jose's drivers". During our analysis, we visually compared the bike counts of the two groups

of stations during the period of August 2013 to August 2015, finding that bike counts in stations based in downtown San Francisco were more volatile than other stations.

Moreover, some stations in that area were found to be highly unpredictable due to the high fluctuation in bike counts. Harry Bridges Plaza Station, which has the highest $MaxAE$, is an example of this type of station. Fig. 5. shows the absolute error and the corresponding actual and RF-model-predicted bike counts at this station during a randomly selected week. In fact, we found that in the future the operator company has planned a coming-soon station very near to Harry Bridges Plaza Station to increase its capacity [48]. As Fig. 5 shows, the fluctuation occurs suddenly in one single observation, during which actual bike counts may rise or fall steeply in a very short period. This high fluctuation in bike counts at these stations can be divided into two types: 1) fluctuation due to the periodic redistribution operation to rebalance bike counts; 2) fluctuation due to the high incoming/outgoing demand in the station within a relatively short period. When we studied the area around Harry Bridges Plaza Station, we hypothesize that this high incoming/outgoing demand comes from it being an open air area at the end of market and restaurants, where artists, skaters, tourists and others congregate to enjoy the happenings and beautiful scenery [49].

It was difficult to classify all the fluctuating incidents as type one or type two. Nonetheless, the first type of fluctuation can be addressed by adding *'rebalancing difference'* to the predicted bike counts ($\hat{y}_{t+\Delta}^i$) if the number of redistributed number and the time of redistribution operation are available. However, this solution is not applicable in all cases.

To further address fluctuation, we then studied the effect of using bike count memory data at station $i$ as a prediction variable by ranging the memory from $t-1 \ to \ t-7$, in which $t$ in Fig. 6. is the model without including any memory data. Results in Fig. 6. show that memory as a prediction variable had a relatively small effect on bike count prediction and could not be used to relax fluctuation in this case study.

This finding suggested that it is crucial to use the available bikes at station-level at time $t$, and, to a smaller extent, the available bikes at its neighboring stations at the same time $t$, as variables to predict the number of available bikes in each station $i$. Although the predictive model still lags by one step, using these variables has a notable influence that limits fluctuation by forcing the predicted bike count model to follow the actual bike count instantaneously when fluctuation occurs in the 15-minute prediction horizon time. This process is shown in Fig. 5.

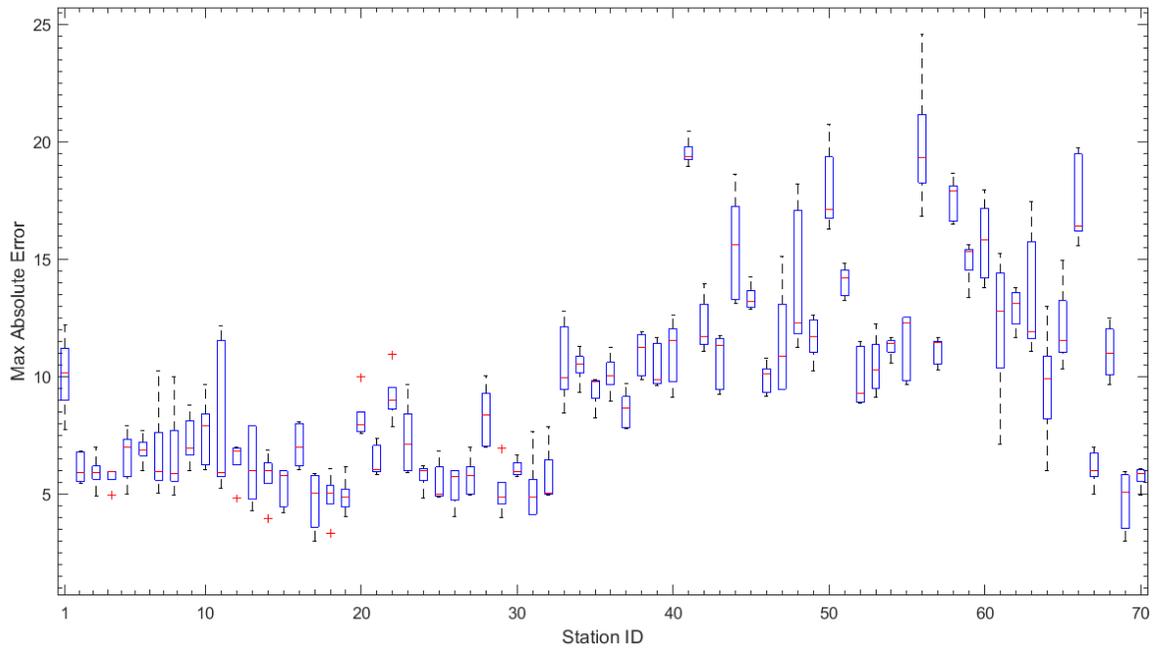

Fig. 4. MaxAE at each station using RF.

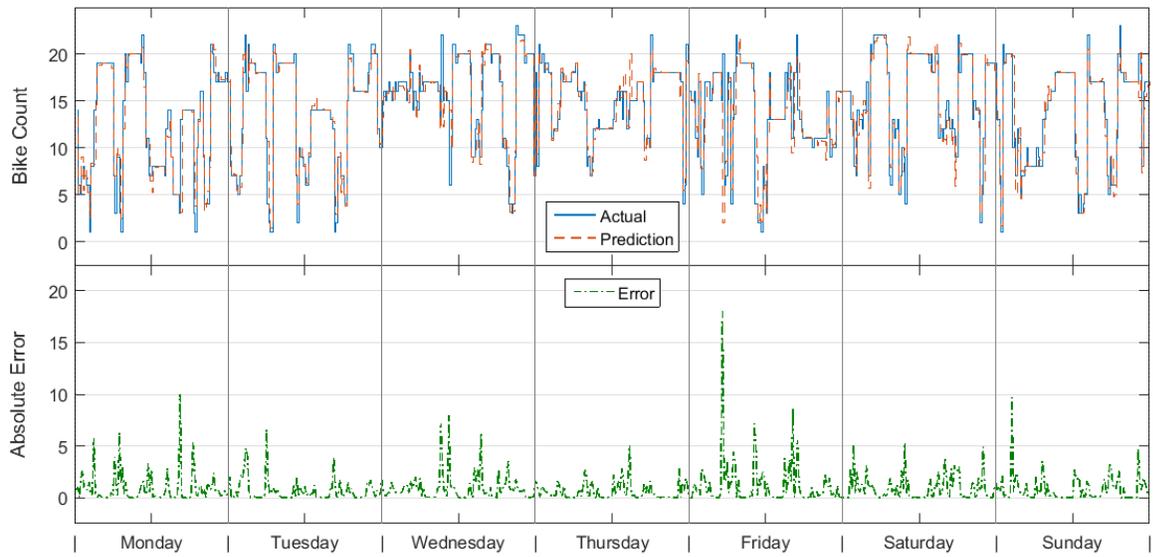

Fig. 5. AE and bike availability at Harry Bridges Plaza Station during a selected week

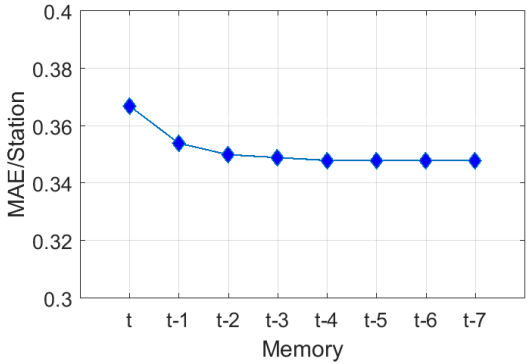

Fig. 6. MAE using RF at different prediction horizons.

**Multivariate Models**

PLSR was used as a multivariate regression to reduce the number of required prediction models for bike stations in the BSS network. When a BSS network has a relatively large number of stations, tracking all the models for each bike station becomes complex and time-consuming. For that reason, we examined the adjacency matrix of the Bay Area BSS network and found that the network can be divided into five regions, as shown in Fig. 7. The regions that resulted from the adjacency matrix were found to consist of bike stations that shared the same ZIP code. This means that the majority of bike trips occurred within the same region and very few trips occurred in more than one region.

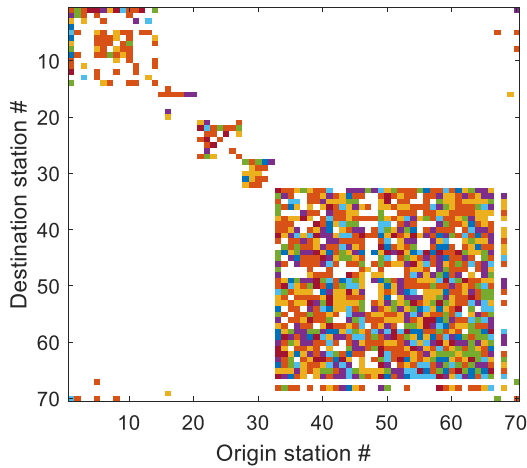

Fig. 7. Adjacency matrix of the Bay Area BSS network.

Using PLSR as a regression algorithm makes it possible to build prediction models considering a multivariate response. Consequently, PLSR was applied to reduce the number of models to five, each of which is specified for one region (i.e., one ZIP code) to reflect the spatial correlation between stations. The input predictors' vector is $X_t^z$, which consists of the available bikes in each region $z$ at time $t$, the month of year, day of week, time of day, and various selected weather conditions. The response's vector is $\log(Y_{t+\Delta}^z)$, where $z = 1, 2, 3, 4, 5$, which is the log of the number of available bikes at all stations in each of the studied regions $z$ at a prediction horizon time $\Delta$ (ranges 15–120 minutes). We found that the prediction errors for PLSR were higher than the RF and LSBoost prediction errors when Δ = 15 minutes, as shown in Fig. 8. Although the prediction errors resulting from PLSR were higher than the previous results, the resulting models from PLSR were sufficient and desirable for relatively large BSS networks.

These results show that the proposed methods outperformed the previously used method to tackle this problem. First, most of the previous studies used time series analysis (e.g. ARIMA and ARMA) to predict bike count in the system as a whole. However, they did not use them to predict bike count prediction at the station-level as this study proposed. Second, time series analysis has some limitations. The data is assumed to be stationary, which means that the series properties are no longer based on the corresponding captured time (must have a constant variance and mean). Further, while data should be univariate in the case of time series analysis (only works on a single variable that depends on the past values), we used RF, LSBoost, and PLSR to model the effect of several variables including the available bikes at station i at time t, the available bikes at its neighbors at the same time t, the month of year, day of week, time of day, and various selected weather information (temperature, visibility, humidity, wind speed, precipitation level, and if it was a rainy, foggy, or sunny day). Modeling all these variables using a time series analysis would make the problem very complicated.

Given the type and the size of the data we've selected RF and LSBoost for the station-level prediction and PLSR for the network-level prediction. On one hand, RF and LSBoost as regression trees offer many advantages that overcome the issues in our data. there are very few or no assumptions when applying it in most of the cases, it is well-suited and robust when there is a risk of overfitting, there are no need to search for data transformations to fit, it is robust against multicollinearity (can handle extremely linear and nonlinear relationships between continuous numerical variables and categorical), it can find the relative *importance* of each variable by ranking their contribution in the predictive model. On the other hand, PLSR offers many advantages making it the best model for our data. It can robustly handle various descriptor variables, as well as nonorthogonal variables that may need multiple responses, while at the same time it still can provide a promising predictive accuracy and it reduces the risk of correlation. Moreover, we used RF and LSBoost for the same type of analysis as we wanted to provide more insights and comparison between the two algorithms.

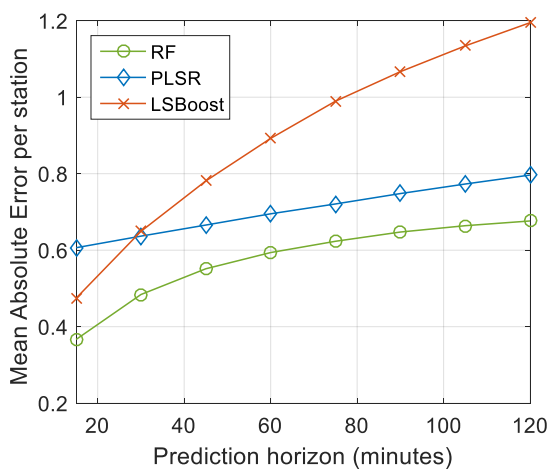

Fig. 8. PLSR, RF, and LSBoost MAE at different prediction horizons.

CONCLUSION

This study proposed modeling bike count at the San Francisco Bay Area BSS at the station-level using machine learning algorithms, which has not been studied well to the best of our knowledge. We used Random Forest and Least-Squares Boosting as univariate regression algorithms to model the number of available bikes at the station-level. We then used a multivariate model (i.e. PLSR) to consider the spatiotemporal correlation between different stations and to facilitate implementing this when applied to relatively large BSS network and examined using a multivariate model to predict the available bikes at the network-level. Results showed that RF slightly outperformed LSBoost with MAE of 0.37 bikes/station and 0.58 bikes/station, respectively. We found that although the univariate models in general slightly performed better than the multivariate model, the multivariate model result of MAE as 0.6 bikes/station might be applicable and practical in the case of predicting bikes on the network-level when the BSS contains relatively many stations.

Investigating BSSs at the station-level is the comprehensive prediction problem that will support planners, operators, and policymakers to effectively make decisions such as considering the moving underused stations (or locating new ones) in the more congested regions in the system. The results from bike counts at stations located in downtown San Francisco were more volatile than the other stations in the BSS network. This high fluctuation in bike counts at these stations can be divided into two types: 1) fluctuation due to the periodic redistribution operation to rebalance bike counts; 2) fluctuation due to the high incoming/outgoing demand at the stations within a relatively short period of time. These findings suggest that demographic information and some environmental factors are significant in modeling bike counts. Although the predictive model still lags by one step, the available bikes modeled at the station-level at time $t$ had a noticeable influence on the prediction of bike counts at $t + \Delta$.

The results of this study provide new real-world insights to policymakers and operators. As we found that the resulted stations in each region of the adjacency matrix from the multivariate analysis happens to be sharing the same ZIP code, this means that most of the trips were short distance trips. This might affect the idea of having overtime fees applied for longer than 30-minute trips. Station neighbors and prediction horizon times were found to be significant predictors, with 15 minutes being the most effective prediction horizon time. Specifically, when the prediction horizon time extends, we found that the corresponding error rises. This is could be helpful to know how to manage BSSs more responsively and efficiently and accomplish better operating functioning.

ACKNOWLEDGMENT

This work is supported in part by the National Science Foundation via grant #DGE-1545362, UrbComp (Urban Computing): Data Science for Modeling, Understanding, and Advancing Urban Populations.

DATA AVAILABILITY

Data used in this study is publicly available at https://www.kaggle.com/benhamner/sf-bay-area-bike-share.